\documentclass[3p,times]{elsarticle}

\usepackage{ecrc}

\volume{00}

\firstpage{1}

\journalname{Nuclear Physics A}

\runauth{C.~E.~Coleman-Smith et al.}

\jid{nupha}

\jnltitlelogo{Nuclear Physics A}

\usepackage{amsmath, amssymb, amsfonts}
\usepackage{graphicx}

\usepackage[figuresright]{rotating}

\newcommand{\figref}[1]{Fig.\,\ref{#1}}

\newcommand{\fm}{\thinspace{\text{fm}}\thinspace}

\begin{document}

\begin{frontmatter}



\dochead{}

\title{What can we learn from dijets? A systematic study with VNI/BMS.}


\author{C.~E.~Coleman-Smith}
\ead{cec24@phy.duke.edu}
\author{B.~M\"uller}
\address{Department of Physics, Duke University, Durham, NC 27708-0305}

\begin{abstract}
We present a systematic study of dijet suppression at RHIC using the VNI/BMS parton cascade code.
We examine the medium modification of the dijet asymmetry $A_j$ and the energy distribution within the dijets. Understanding the sensitivity of these observables to properties of the deconfined medium and to
experimental factors is vital if dijets are to be useful for QCD tomography. VNI/BMS provides a controllable
test-bed with sufficient complexity to model jet modification without confounding results with fluctuations
from hydrodynamics and hadronization. Dijets are examined under the modification of: the jet transport
coefficient $\hat{q}$; the path length of leading and sub-leading jets; jet cone angle and the 
jet-medium interaction mechanism. We find that $A_j$ is very sensitive to the distance traveled by
the secondary jet, the jet shape is dominated by $\hat{q}$ and the nature of the interaction mechanism.
\end{abstract}


\end{frontmatter}



\section{Introduction}

The suppression of high energy  $E_{t} \sim 100-200$~GeV dijets in heavy ion collisions has recently been observed at the LHC \cite{Aad:2010bu, Chatrchyan:2011sx}. These results have shown the feasibility of using dijets as correlated probes of jet modification in hot QCD matter. Proposed upgrades to PHENIX \cite{Aidala:2012nz} along with the continuing development of jet reconstruction at STAR will allow access to dijets in the RHIC kinematic window $E_{t} \sim 15-65$~GeV.  The dijet asymmetry is defined as \cite{Aad:2010bu}
\begin{equation}
  \label{eqn-aj-define}
A_j = \frac{E_{t,\ell} - E_{t,s}}{E_{t,\ell}+E_{t,s}},
\end{equation}
where $E_{t,\ell}$ is the transverse energy of the leading jet and $E_{t,s}$ is that of the sub-leading jet. The asymmetry distribution measured in Pb+Pb collisions at the LHC has been successfully reproduced with a variety of models including the one considered here \cite{Qin:2010mn,Young:2011qx,ColemanSmith:2011wd,He:2011pd}. Recent work by Renk \cite{Renk:2012cx} indicates that these results may be relatively insensitive to the fine details of jet energy loss, suggesting that $A_j$ may be unsuitable for tomographic purposes. If this is indeed the case, are there other more differential dijet observables for jet measurements at RHIC? 

With these issues in mind we have used the VNI/BMS parton cascade to  systematically explore dijet suppression at RHIC energy scales under controlled medium conditions. We examine the modification of dijets under variation of the medium radius and temperature, the strong coupling constant and the jet definition in terms of the Anti-Kt cone angle. The VNI/BMS parton cascade model code \cite{Geiger:1991nj,Bass:2002fh} provides access to the full jet/medium development at a fixed $\hat{q}$. We run the code in a static uniform-medium mode. The medium is modeled as a torus of a given radius. The dijet propagation lengths are generated as chords centered on uniformly sampled hard-collision vertices within this torus. The model includes a partonic medium which is treated on an equal footing with the jet. This allows jet partons to escape into the medium and vice-verso. All results are presented at the partonic level only. We use anti-kt jet reconstruction \cite{Cacciari:2008gp} throughout to give a somewhat realistic treatment of the jet measurement process.

\section{The Parton Cascade} 

The parton cascade model (PCM) is a Monte-Carlo implementation of the relativistic Boltzmann transport of quarks and gluons
\begin{equation}
\label{eqn-boltzmann}
p^{\mu} \frac{\partial}{\partial x^{\mu}} F_k(x, p) = \sum_{i}\mathcal{C}_i F_k(x,p).
\end{equation}
The collision term $\mathcal{C}_i$ includes all possible $2\to2$ interactions and final-state radiation $1 \to n$
\begin{align}
  \label{eqn-pcm-collision}
  \mathcal{C}_i F_k(x,\vec{p}) &= \frac{(2 \pi)^4}{2 S_i} \int \prod_{j} d\Gamma_j | \mathcal{M}_i | ^2 \,\delta^4\left(P_{\rm in} - P_{\rm out}\right) D(F_k(x, \vec{p})),
\end{align}
$d\Gamma_j$ is the Lorentz invariant phase space for the process $j$, $D$ is the collision flux factor and $S_i$ is a process dependent normalization factor. A geometric interpretation of the total cross-section is used to select pairs of partons for interaction. Between collisions, the partons propagate along straight line trajectories. 

In the VNI/BMS implementation of the PCM outgoing off-shell partons are brought back on-shell through a medium modified time-like branching. The partons created in this process are subject to a Monte-Carlo LPM effect \cite{Zapp:2008af, ColemanSmith:2011wd}. The strong coupling constant for scatterings is held fixed at $\alpha_s = 0.3$ although we shall explore the effects of its variation below. A quark-gluon plasma is simulated as a box of thermal quarks and gluons generated at some fixed temperature. Periodic boundary conditions are imposed on the box, whose size is selected to be large enough that if a simulated jet wraps around it will not interact with its own wake. 

The partonic contents of jets, created by the event generator PYTHIA-8 \cite{Sjostrand:2006za, Sjostrand08pythia8} and evolved down to $Q_0 = 1$~GeV, are injected into the medium. Each sub-jet is evolved separately. The jet evolution is recorded without reference to a particular jet definition (cone-angle, energy cuts, jet reconstruction algorithm, etc). Each particle in the initially inserted jet is marked as being ``jetty'', this jettiness tag is then iteratively applied to all partons that interact with an already jetty-labeled parton. The entire time-evolution of the jet can then be reconstructed using a suitable jet finder (FastJet \cite{Cacciari:2005hq}) and jet definition. In this way the influence of varying jet definitions can be readily studied. This process factorizes the initial vacuum radiation process from the medium induced radiation, this is a weakness that will be addressed in future work.

VNI/BMS is a \emph{simple enough} jet-suppression model. While it may not be suitable for use as a full event generator, lacking hydrodynamical flow and hadronization, the active medium and medium-modified radiation make it a useful test bed for many but not all jet modification features.

\section{Results}

We generate candidate Monte-Carlo dijet events from pp collisions at $\sqrt{s} = 200$~GeV, anti-kt reconstruction was used to identify satisfactory dijet events \cite{Cacciari:2008gp}. The following kinematic requirements were always imposed: we fix the minimum momentum of an identified jet to be $p_{t,\rm min} = 5$ GeV, $\Delta |\eta_{12} | < 1.1$, $\Delta \phi_{12} > \pi / 2$ where $\Delta \eta_{12}$ and $\Delta \phi_{12}$ are measured between the two reconstructed jets in the event. The constituent partons of each identified jet are translated from PYTHIA into the parton cascade medium for modification. We consider the dijet asymmetry $A_j$ and the partonic jet profile.

\begin{figure}[ht]
  \begin{minipage}[b]{0.45\linewidth}
  \centering
  \includegraphics[width=0.75\textwidth, clip, trim=0.2cm 0.1cm 0.5cm 0.1cm]{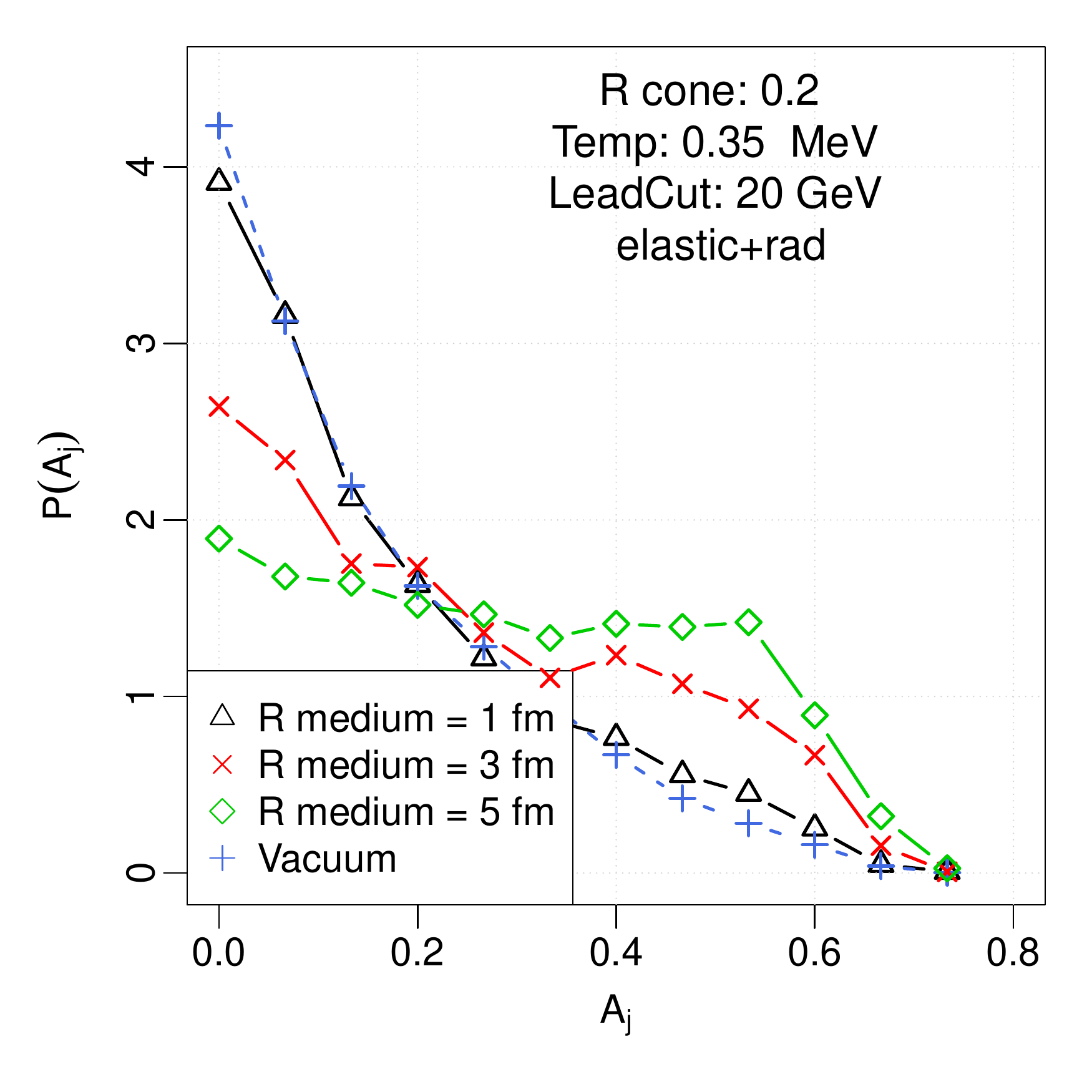}
  \caption{ The dijet asymmetry as a function of the path length of the leading and sub-leading jets,  for a medium with $T=350$~MeV.} 
  \label{fig-aj-z}
  \end{minipage}
    \hspace{0.5cm}
  \begin{minipage}[b]{0.45\linewidth}
    \centering
  \includegraphics[width=0.75\textwidth, clip, trim=0.2cm 0.1cm 0.5cm 1cm]{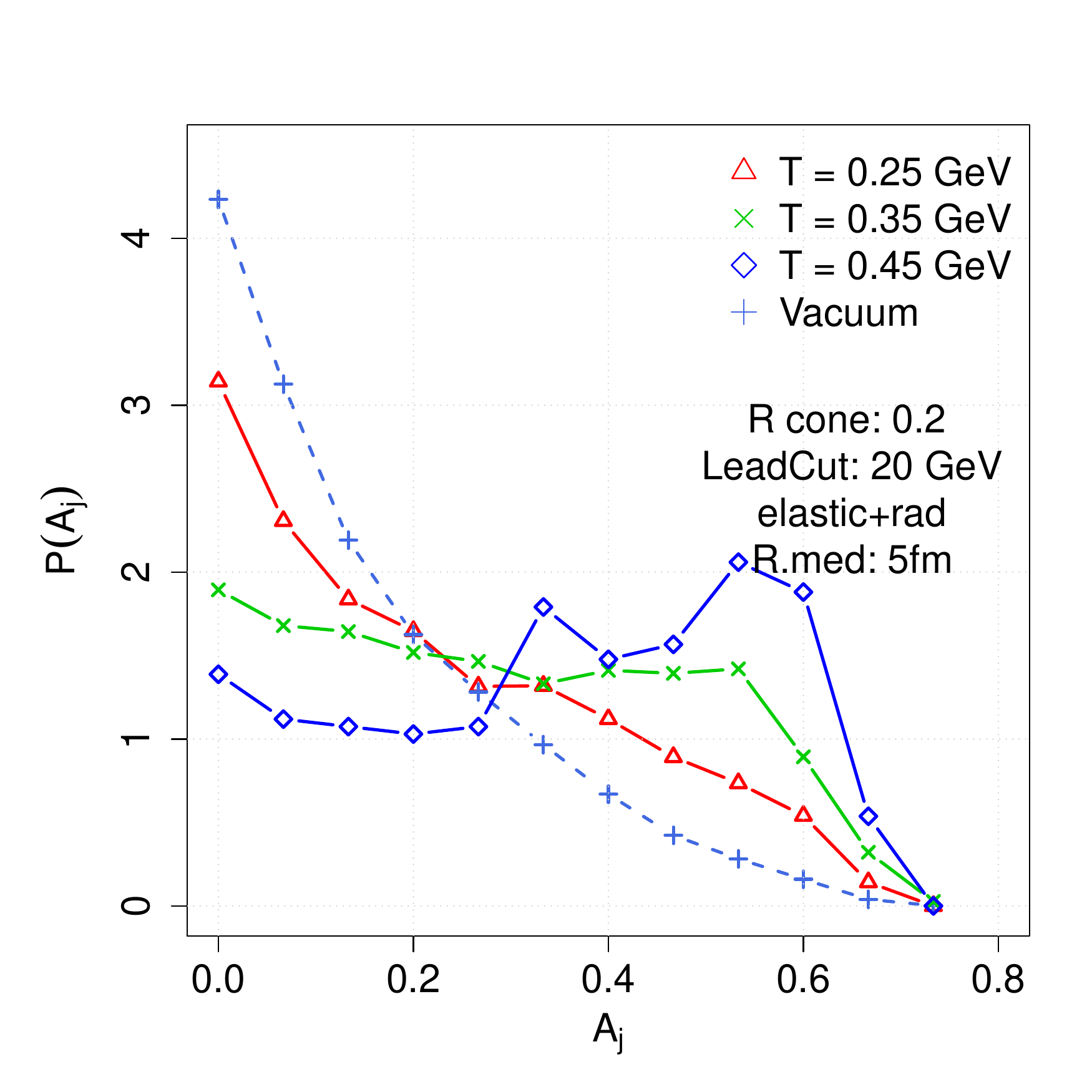}
  \caption{The influence of the medium temperature on the asymmetry of dijets with $E_{t1} \ge 20$ GeV.}
  \label{fig-aj-temp}
  \end{minipage}
\end{figure}

In \figref{fig-aj-z} we show the variation of $A_j$ with the medium radius. The vacuum jet distribution falls off very quickly, jets with an initial $E_{t,\ell} > 65$~GeV are extremely rare. Increasing the medium temperature from $250$~MeV to $350$~MeV (see \figref{fig-aj-temp})  leads to a depletion of jets with a small modification, shifting the dijet distribution towards higher $A_j$ values.  By applying an energy cut to differentiate the leading and sub-leading jets in a dijet pair we have biased our results. We select for leading jets which travel a short distance, and are less modified,  along with sub-leading jets that travel a very long distance and so experience a larger integrated $\hat{q}$. This surface bias gives a large contribution to $A_j$ relative to the vacuum result. For the remainder of the analysis we have fixed $R_{med} = 5$~\fm.

Since $\hat{q} \propto T^3$ the dijet asymmetry depends strongly upon the medium temperature. In \figref{fig-aj-temp} we show results of varying the medium temperature. Increasing medium temperature and therefore $\hat{q}$ leads to increased jet-medium interactions and a strong swing in the observed $A_j$. 
\begin{figure}[ht]
  \begin{minipage}[b]{0.45\linewidth}
  \centering
  \includegraphics[width=0.75\textwidth]{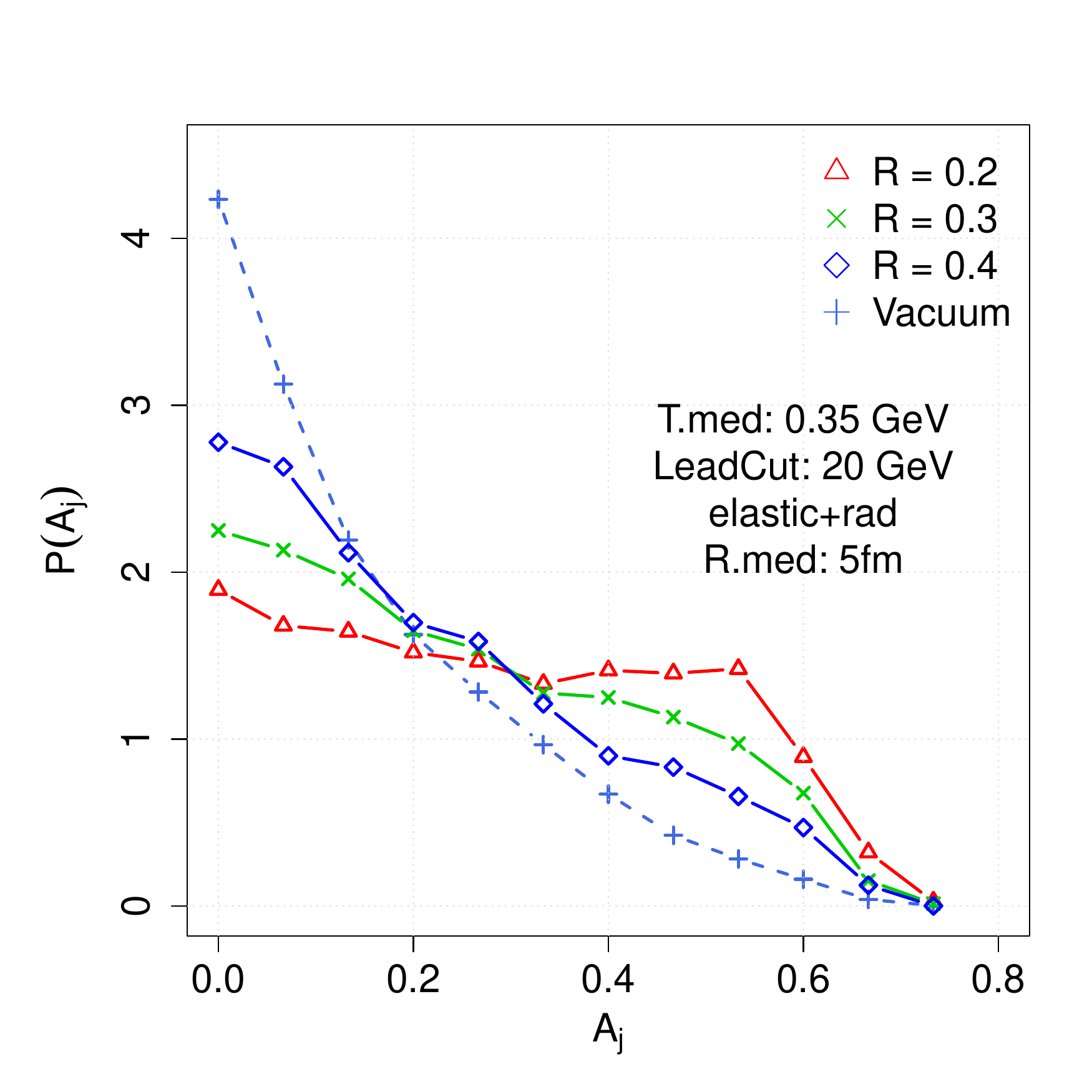}
  \caption{The variation of $A_j$ with  $R$ the anti-kt jet cone radius. The medium radius is fixed at $R_{med} = 5$~\fm with $T=0.35$~GeV and $E_{t,\ell} > 20$ GeV. }
  \label{fig-aj-R}
  \end{minipage}
  \hspace{0.5cm}
  \begin{minipage}[b]{0.45\linewidth}
  \centering
  \includegraphics[width=0.75\textwidth, clip]{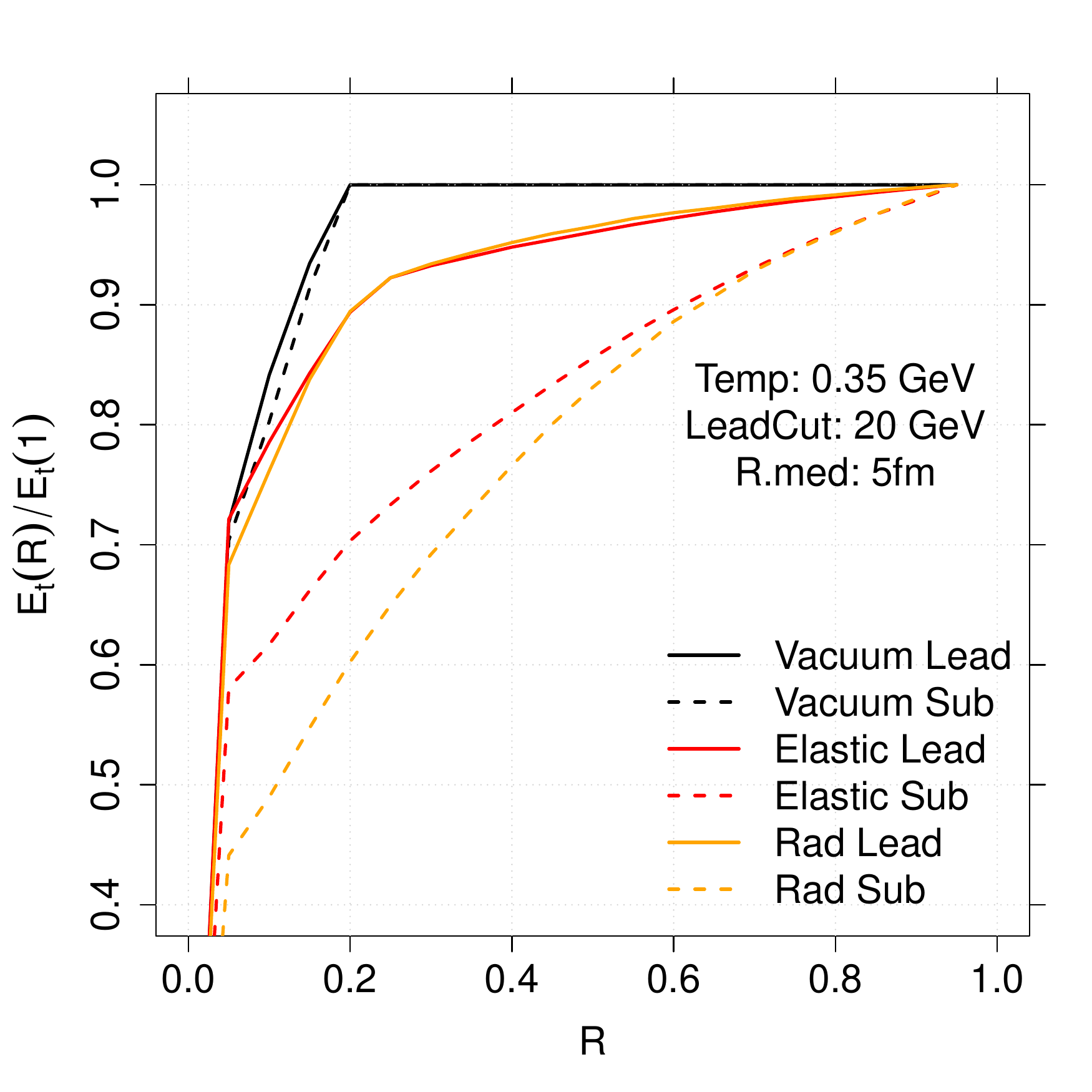}
  \caption{Variation of the jet radial profile at $T=350$~MeV. Leading jet profiles are shown as solid lines and sub-leading jets are dashed. Results for elastic energy loss only are shown in red; those including radiative and elastic energy loss are in orange.}
  \label{fig-js-tvary}
  \end{minipage}
  \hspace{0.5cm}
\end{figure}

In \figref{fig-aj-R} we show $A_j$ as a function of the anti-kt cone angle $R$ for jets in a medium with $T=350$~MeV. As $R$ is increased the amount of dijet modification is significantly reduced. It is important to note that the only partons which can be included as part of the measured jets were either directly created by the jet or are those which have scattered with jet partons. The thermal medium is currently artificially excluded from the jet finder, this removes uncertainty associated with background removal. As $R$ increases more of the relatively soft radiated partons and forward scattered medium partons are included in the jet definition along with original hard core. This leads to higher reconstructed jet energies at larger $R$'s which in turn leads to the observed reduced $A_j$. The effect is proportional to the distance traveled by the sub-leading jet, jets which have traveled shorter distances have built up less of a cloud of soft partons and picked up fewer medium partons by elastic forward scattering. The rate of transverse diffusion of these soft partons is proportional to $\hat{q}$, so the jet-cone/cloud will be wider at higher medium temperatures. 

Let us now examine the modification of the radial jet profile under the same set of factors. We define the radial jet profile as the ratio of jet energy reconstructed within a certain jet-cone radius $R$ relative to the reconstructed energy at $R=1$. This gives a normalized radial profile which is very sensitive to variations in the medium temperature, the jet interaction mechanism and the strong coupling. This observable could be experimentally obtained by a similar process of iterative jet reconstruction. In \figref{fig-js-tvary} we show the radial profile for both elastic and radiatively modified jets. The leading jet profiles (solid lines) are somewhat modified compared to the vacuum jets (black profiles), we see very little separation between the radiative and elastic only leading jet profiles. The sub-leading jet profiles are dramatically modified compared to the vacuum and leading jet profiles. The elastic and radiative profiles clearly separate, the radiative sub-leading jets become broader and softer than the elastic only. Both sets of sub-leading jets become much broader and softer compared to the leading jets.

\section{Conclusions}

We have examined the modification of RHIC scale dijets by a pQCD medium over a range of temperatures. We find that the dijet asymmetry is quite sensitive to the amount of medium encountered by the jets, and very sensitive to variations in $\hat{q}$. The leading jets show a strong surface bias and lose very little energy making their $E_t$ a reasonable proxy for their initial energy. As such $A_j$ appears to give a reasonable measure of the modification of sub-leading jets at these scales.

The jet profiles we present show the strong modification of the sub-leading jets as a function of temperature and strong coupling. They also differentiate between interaction mechanisms, understanding the nature of the jet-medium interaction as well as its strength is of paramount import. Although these profiles may well be sensitive to the non-trivial hadronization process, we believe that they could provide a very informative window into the real nature of jet interactions with the QGP. 

Dijets at RHIC scales are likely to be strongly modified by the presence of the deconfined QGP medium. The observables we have discussed are sensitive to many aspects of this modification and suggest that further jet measurements at RHIC will provide valuable insights into the nature of the QGP and into the applicability of pQCD jet suppression models .





\bibliographystyle{elsarticle-num}
\bibliography{ccs-hp.bib}







\end{document}